\documentclass[12pt]{article}
\usepackage{amssymb, euscript}
\usepackage{amsmath,bbm}

\usepackage{theorem}
\theoremheaderfont{\scshape} 
\theoremstyle{plain}

\newtheorem{example}{Example}

\newcommand{\haken}{\mathbin{\hbox to 8pt{%
                 \vrule height0.4pt width7pt depth0pt
                 \kern-.4pt
                 \vrule height4pt width0.4pt depth0pt\hss}}}

\newcommand{\be}[3]{\begin{equation}  \label{#1#2#3}}
\newcommand{\bea}[3]{\begin{eqnarray}  \label{#1#2#3}}
\newcommand{\ee}{\end{equation}}
\newcommand{\ba}{\begin{array}}
\newcommand{\ea}{\end{array}}
\newcommand{\eea}{\end{eqnarray}}

\newcommand{\M}{ {\cal M} }

\def\genc{\mathop{{\cal J}_{J}}\nolimits}
\def\gens{\mathop{{\cal J}_{\omega}}\nolimits}
\def\gencm{\mathop{\hat{\cal J}_{J}}\nolimits}
\def\gensm{\mathop{\hat{\cal J}_{\omega}}\nolimits}
\def\gencb{\mathop{{\cal J}^B_{J}}\nolimits}
\def\gensb{\mathop{{\cal J}^B_{\omega}}\nolimits}

\renewcommand{\arraystretch}{1.2}

\begin{document}

\baselineskip=20pt
\parskip=6pt


\thispagestyle{empty}

\begin{flushright}
\hfill{HU-EP-04/30} \\
\hfill{hep-th/0406046}

\end{flushright}

\vspace{10pt}

\begin{center}{ \LARGE{\bf
Generalized Calabi-Yau structures \\[4mm]
and mirror symmetry
}}

\vspace{35pt}

{\bf Claus Jeschek}$^a$

\vspace{15pt}

$^a$ {\it  Humboldt Universit\"at zu Berlin,\\
Institut f\"ur Physik,\\
Newtonstrasse 15, 12489 Berlin, Germany,\\
E-mail: jeschek@physik.hu-berlin.de.}\\[1mm]

\vspace{8pt}

\vspace{40pt}

{\bf ABSTRACT}

\end{center}

\noindent
We use the differential geometrical framework of generalized (almost)
Calabi-Yau structures to reconsider the concept of mirror symmetry. 
It is shown that not only the metric and B-field but also the algebraic 
structures are uniquely mapped. 
As an example we use the six-torus as a
trivial generalized Calabi-Yau 6-fold and an appropriate B-field.

\vfill

\newpage


\section{Introduction}


For a long time it has been believed that compactifications of string theory
on compact six-dimensional spaces lead to a satisfactory model 
describing several important features of the real 4-dimensional world.
Applying supersymmetry arguments to vacuum structure without fluxes,
lead to internal manifolds having special holonomy. When one considers 
in addition to the metric, the $B$-field and the dilaton,  
the backgrounds can be characterized by the 
theory of $G$-structures. 

The problem of having several, distinct
but physically relevant superstring theories formulated in ${\mathbb M}^{1,9}$
was solved by duality maps. 
Moreover, using e.g. SCFT, it turned out that (in the flux free case) in 
particular
the low-energy effective theories for IIA and IIB  can be (mirror)dual, 
although the compactified background spaces are even topologically 
inequivalent. By using moduli space investigations of D-branes
it was conjectured in \cite{140}   
that the low-energy effective theories are mirror symmetric to each other 
if the Calabi Yau spaces are $T^3$ fibered and the action of T-duality on to
these fibres is the mirror map. This means that firstly only a very restricted
subspace of the huge moduli space of Calabi-Yau spaces is relevant 
(being compatible with the duality). Also considering the remaining 
(background) fields, no satisfactory mathematical formalism
to combine the concept of $G$-structures and duality maps in a natural way
has been found yet.

We consider this problem by using the concept of 
generalized complex spaces introduced by Hitchin in \cite{185} and further 
developed by Gualtieri\cite{190}. On the one hand, this concept provides 
us with a natural map, called the $B$-field transformation, where 
$B\in\Lambda^2T^{\ast}$
\footnote{It is convenient to consider this $B$-field in 
the following as the physical background $B$-field}. 
On the other hand, we introduce a well-defined map ${\mathcal M}$ 
relating generalized Calabi-Yau spaces to each other which we conjecture 
therefore as a mirror map. 
 
After reviewing the necessary definitions in section two, we apply the theory
of generalized Calabi-Yau manifolds (for convenience only) to the simple 
example of $T^6$ in section three. Here we investigate the case  
without a $B$-field first. From a theory inherited 
$B$-field transformation, we introduce a $B$-field in a systematic way. 
Moreover, we give an explicit description 
of the mirror symmetry map and derive the mirrored generalized 
structures where we are able to rederive the Buscher rules as well
as the mirrored (classical) complex and symplectic structures. 
In section four we discuss obvious generalizations.

Recent developments using generalized complex structures and related topics
can be found e.g. in \cite{110,150,160,180,185,190,200,210,220,230,240}.

{\bf Note added.} While preparing this paper we became aware of the
independent work done by Oren Ben-Bassat \cite{210} which has some overlap
with our results.

\section{Preliminaries}
%
The theory of generalized complex (and Calabi-Yau) structures was introduced 
by Hitchin \cite{185}.
Recently, Gualtieri \cite{190} introduced in his thesis the notation
of generalized K\"ahler structures where he discussed also integrability
conditions and torsion. In what follows we will stick (almost) to the
definitions given there and find it usefull to remember here the relevant
concepts.
%
\subsection{Basic definitions}
Let $T$ be a 6-dim real vector space and $T^{\ast}$ its dual. Because of
treating manifolds we also use (in abuse of notation) the same symbols
for the (co-)tangent bundle. By introducing local coordinates and using 
the canonical basis for $T$ and $T^{\ast}$ we also have the natural pairings:
\be100
dx^{\mu}(\partial_{\nu}) = \delta^{\mu}{}_{\nu}\, , \qquad
\partial_{\mu}(dx^{\nu}) = \delta_{\mu}{}^{\nu}
\ee
Using this fact we define the non-degenerate, symmetric bilinear form of 
signature $(6,6)$ on the vector space $T\oplus T^{\ast}$ by
\be105
\langle X + \xi , Y + \eta \rangle = \frac{1}{2}(\xi(Y)+\eta(X))
\ee
The group preserving this bilinear form and the orientation is the non-compact
special orthogonal group $SO(6,6)$. The Lie algebra $\mathfrak{so}(6,6)$ can
be decomposed in a direct sum of three terms. An element 
$g\in \mathfrak{so}(6,6)$ is given by
\be106
g=\begin{pmatrix} A & \beta \\ B & -A^{\ast} \end{pmatrix}
\ee
Exponentiation gives three distinguished transformations: 
the diagonal embedded \\
$GL(T)$ action and the two shear transformations generated by
$B\in\Lambda^2T^{\ast}$ and $\beta\in\Lambda^2T$.
The $B$-transformation is given by
\be107
\mbox{exp}(B)=\begin{pmatrix} 1 &  \\ B & 1 \end{pmatrix}
\ee
which means that: $\mbox{exp}(B)(X + \xi) = X + \xi + X\haken B$.
Correspondingly for the $\beta$-transformation we obtain
\be108
\mbox{exp}(\beta)=\begin{pmatrix} 1 & \beta \\  & 1 \end{pmatrix}
\ee
where: $\mbox{exp}(\beta)(X + \xi) = X + \xi + \xi\haken \beta$.
\subsection{Spinors and associated bilinear form}
Let us act with $X + \xi\in T\oplus T^{\ast}$ on 
$\varphi\in\Lambda T^{\ast}$ by
\be120
(X+\xi)\cdot\varphi=X\haken\varphi + \xi\wedge\varphi
\ee
and note that $(X+\xi)^2\cdot\varphi=\langle X+\xi \rangle \varphi$,
where we used (\ref{105}).
This defines the spin representation on the exterior algebra
$\Lambda T^{\ast}$. Taking the argument of dimension and signature into
account the spin representation splits into two chiral irreducible parts
$S= S^+ \oplus S^-$. Furthermore, we consider elements of $S^+(S^-)$ as 
even(odd) forms. In what follows we prefer the notation of ``even''(``odd'')
forms instead of using $S^{\pm}$.
Let us define on the spinor bundles $S^{\pm}$ an invariant bilinear form 
\be150
(\, \cdot \, , \, \cdot\, ): \, S \otimes S \to \mbox{det} \,  T^{\ast}.
\ee
Because of dealing here with manifolds of real dimension $n=6$ 
the bilinear form is skew-symmetric: 
\be160
\langle \varphi, \psi \rangle = (\sigma\varphi\wedge\psi)_{top}
\ee
where we used the anti-automorphism
\be170
\sigma(\varphi_{2m})=(-1)^m\varphi_{2m}\, ,\qquad
   \sigma(\varphi_{2m+1})=(-1)^m\varphi_{2m+1}\,.
\ee
\subsection{Purity}
In this subsection we introduce (less familiar for physicists) the powerfull
tool of pure spinors. Let $L_{\varphi}\subset T\oplus T^{\ast}$ be defined
by using the clifford multiplication such that
\be200
L_{\varphi} = \{ X + \xi \in T\oplus T^{\ast} | (X+\xi)\cdot\varphi=0 \}
\ee
It can be easily checked that $L_{\varphi}$ is isotropic. Spinors with
maximally isotropic associated annihilator $L_{\varphi}$ (or null space) 
are called pure, i.e. $\varphi$ is pure when $\mbox{dim}(L_{\varphi})=6$.
The power of pure spinors come into play by using the fact that: Every
maximal isotropic subspace of $T\oplus T^{\ast}$ is generated by a unique
pure spinor line. Using the bilinear form on the spinor bundle we can 
distinguish two maximal isotropics $L_{\varphi},L_{\psi}$:
\be210
L_{\varphi}\cap L_{\psi}=0 \qquad \Leftrightarrow \qquad
         0\neq\langle \varphi, \psi \rangle =(\sigma\varphi\wedge\psi)_{top}
\ee
where $\varphi,\psi$ are pure spinors.
Note that every maximal isotropic subspace $L$ of type $k$ has the form
\be215
L(E,\varepsilon)=\{ X+\xi \in E\oplus T^{\ast} \, : \xi|_E = \varepsilon(X)\}
\ee
where $E\subset T, \varepsilon\in \Lambda^2 T^{\ast}$ and $k$ is the
codimension of its projection onto $T$. In what follows we only consider
the special case when $\varepsilon=B$ and only introduce the extremal
isotropics of lowest and highest type. 
It is possible to extend our previous facts by complexification to
$(T\oplus T^{\ast})\otimes {\mathbb C}$. This provides us also with the
action of complex conjugation. 
We are now prepared to define one of the important working tools by which
we construct later on generalized Calabi-Yau structures:
The complex maximal isotropic subspace 
$L\subset (T\oplus T^{\ast})\otimes {\mathbb C}$ ($E$ denotes its projection
on  $T\otimes {\mathbb C}$ with $\mbox{dim}_{ {\mathbb C}}(E)=3-k$)
is defined by the complex spinor line 
$U_L\subset\Lambda(T^{\ast}\otimes {\mathbb C})$ which is generated by
\be220
\varphi_L=c\cdot \mbox{exp}(B+i\,\omega)\theta_1\wedge\ldots\wedge\theta_k,
\ee
where 
$c\in{\mathbb C}, (B+i\,\omega)\in \Lambda^2(T^{\ast}\otimes {\mathbb C})$
and $\theta_i$ are linearly independent complex one-forms.
Note: If $\theta_1\wedge\ldots\wedge\theta_k$ is pure then one can easily check
that clifford multiplication by $\mbox{exp}(B+i\,\omega)$ keeps the
property of purity. But obviously we may single out different isotropics.
The reader might now wonder if the real form $\omega\in \Lambda^2 T^{\ast}$
can be the symplectic form. The answer in our case is 'yes'. Later on we
define two spinors, one is of type $k=0$ (symplectic type) and one is of
type $k=3$ (complex type). We do not need in this note non-extremal types.
%
\subsection{Integrability}
%
In the ordinary sense (at least in diffenential geometry) we call a 
structure integrable if smooth vector 
fields are closed under the Lie bracket. The structure of the Lie bracket,
however, is invariant only under diffeomorphisms. The situation changes
if we ask for integrability of smooth sections of 
$(T\oplus T^{\ast})\otimes {\mathbb C}$. The answer of this question is
the Courant bracket. Our main concern are smooth sections of maximal 
isotropic sub-bundles which are closed under this bracket. 
It is also shown that the Courant bracket is additionally invariant 
under $B$-transformations iff $dB=0$. 
Clifford multiplication of $X + \xi \in T\oplus T^{\ast}$ on a spinor is
a map taking $\Lambda^{ev/od}\to\Lambda^{od/ev}$ (see (\ref{120})). 
Also the exterior derivative is such a map.
There is the following correspondence between isotropic $L$ being 
involutive (closed under Courant bracket) and smooth sections in the spin 
bundle:
\be250
L_{\rho}\mbox{ is involutive} \qquad \Leftrightarrow \qquad
  \exists (X+ \xi)\in C^{\infty}(T\oplus T^{\ast})\otimes {\mathbb C}:
        \, d\rho = (X+ \xi)\cdot\rho
\ee
for any local trivialization $\rho$. 
We can also extend this definition by twisting the Courant bracket with a 
gerbe. Integrability forces the substitution of the ordinary differential
operator $d$ by the twisted differential operator $d^H$:
\be255
d\, \cdot \to d^H \, \cdot=d\, \cdot +H\wedge\cdot
\ee
where $H\in\Lambda^3 T^{\ast}$ is real and closed.
%
\subsection{Generalized Calabi-Yau structures}
%
Consider the natural indefinite metric (\ref{105}) on $T\oplus T^{\ast}$. 
In what follows we are interested not only in involutive but additionally 
also in positive (negative) definite 
subspaces/subbundles, called $C_{\pm}$. This forces the structure group
to reduce globally to the maximal compact subgroup $O(6)\times O(6)$. We
get therefore the splitting $T\oplus T^{\ast}=C_+\oplus C_-$. This serves
us to define the positive definite metric $G$ on $T\oplus T^{\ast}$.
The metric $G$ has the properties of being symmetric 
($G^{\ast}=G$) and squares to one ($G^2=1$). 
(Note that $G$ is an automorphism).

A generalized complex structure is an endomorphism $\cal J$ on 
$T\oplus T^{\ast}$ which commutes (is compatible) with $G$. So $C_{\pm}$ 
is stable under the action of $\cal J$. It satisfies 
${\cal J}^2=-1$ and its dual $\cal J^{\ast}$ 
is symplectic ($\cal J^{\ast}=-\cal J)$. Using the properties of $G$ and 
$\cal J$ we can define another generalized complex structure. Additionally, 
requiring that the two commuting generalized complex structures are integrable
we have a generalized K\"ahler structure and $G$ is given by
\be260
G = -{\cal J}_1 {\cal J}_2.
\ee
This reduces the structure group to $U(3)\times U(3)$.

A generalized Calabi Yau structure is a generalized K\"ahler structure
with the following additional constraint for the generating spinor lines,
\be265
(\sigma\varphi_1\wedge\bar\varphi_1)_{top}= c \cdot \,
   (\sigma\varphi_2\wedge\bar\varphi_2)_{top} 
              \qquad \mbox{on each point}
\ee
where $c\in{\mathbb R}$\,. The property 
\be270
(\sigma\varphi_{1/2}\wedge\bar\varphi_{1/2})_{top} \neq 0
\ee
trivializes the determinant bundle of the maximal isotropic and thus
reduces the structure group to $SU(3)\times SU(3)$.
%
\section{Application}
%
Let us use the powerfull machinery introduced in the previous section to
reproduce commonly known facts. Getting started, we choose the six-torus
$T^6$ and the
complex structure $J$, symplectic structure $\omega$ and metric $g$. We
consider this manifold as a trivial fibration of $T^3\hookrightarrow T^6$
over the base space ${\cal B}=T^3$. Later on we investigate
manifolds having also non-trivial $T^3$ fibrations (also a
non-vanishing $B$-field). The basic idea is to embed the given structures
into generalized ones and consider their behavior under a special map, 
the mirror symmetry map ${\cal M}$.
%
\subsection{Warm up: $T^6$ without $B$-field}
%
We start with $T^6$ and vanishing $B$-field. For later convenience we denote 
the usual structures by $(2\times 2)$-blocks,
strictly speaking, these denote the coordinate matrix of the considered tensor
respecting the base-fibre split of coordinates.
In local coordinates we have
\be300
g= \begin{pmatrix} \delta_{ij} &  \\ 
                 & \delta_{\alpha\beta}\end{pmatrix},
 \qquad
g^{-1}=\begin{pmatrix} \delta^{ij} &  \\ 
                  & \delta^{\alpha\beta}\end{pmatrix}
\ee
where $(y_i,x_{\alpha})$ ($(i,\alpha)\in\{1,2,3\}$) denotes the coordinates 
on the base ${\cal B}$ and fibre ${\cal F}$, respectively. 

The complex and symplectic structures we use are therefore given by
\be305
J= \begin{pmatrix}  & -\delta_{\alpha}{}^i \\ 
               \delta_i{}^{\alpha} & \end{pmatrix},
 \qquad
J^T=\begin{pmatrix}  & \delta^{\alpha}{}_i \\ 
               -\delta^i{}_{\alpha} & \end{pmatrix}
\ee
and
\be306
\omega=\begin{pmatrix}  & -\delta_{\alpha i} \\ 
               \delta_{i\alpha} & \end{pmatrix},
 \qquad
\omega^{-1}=\begin{pmatrix}  & \delta^{\alpha i} \\ 
               -\delta^{i\alpha} & \end{pmatrix}.
\ee
It is not difficult to check that the identities
\be307
J\omega^{-1}=g^{-1} \qquad\mbox{and}\qquad J^T\omega=g
\ee
hold.
Let us embed these structures into the generalized
structures ${\cal J}_{J},{\cal J}_{\omega}$ by:
\be310
{\cal J}_{J}=\begin{pmatrix} J &  \\ 
                & -J^T \end{pmatrix},
 \qquad
{\cal J}_{\omega}=\begin{pmatrix}  & -\omega^{-1} \\ 
               \omega & \end{pmatrix},
\ee
which makes ${\cal J}_{J},{\cal J}_{\omega}$ into generalized complex 
structures. 
Using the above identities (\ref{307}) we can prove easily the identity: 
\be315
G = -{\cal J}_{J}{\cal J}_{\omega}=
\begin{pmatrix}  & g^{-1} \\ 
               g & \end{pmatrix}.
\ee
The triple $({\cal J}_{J},{\cal J}_{\omega},G)$ provides us with a simple
example of a generalized K\"ahler structure. The generating spinor
lines are given by
\be320
\varphi_1=e^{i\, \omega} \Leftrightarrow {\cal J}_{\omega}, \qquad\qquad
\varphi_2=\Omega^{(3,0)} \Leftrightarrow {\cal J}_{J} \,.
\ee
Naturally, by introducing a trivialization and squaring the 
generating spinor lines according to (\ref{265}) we get global 
non-vanishing sections of the canonical bundle and furthermore 
\be325
\omega^3=\frac{i\,3!}{2^3}\,\Omega\wedge\bar\Omega
\ee
where $c=1$. The sections are closed and thus reduces the 
structure group to $SU(3)\times SU(3)$, but obviously the simple example
of $T^6$ has structure group the identity, thus, a trivial generalized 
Calabi-Yau space.

This fixes our setup in the generalized sense and we are now prepared to
define a map ${\cal M}$ which assigns to the generalized Calabi-Yau 
structure an
other generalized Calabi-Yau structure. By specializing only on spaces which
are $T^3$ fibrations over the base ${\cal B}$ and acting only on the fibre
we will call this map ${\cal M}$ a mirror symmetry map. This map is an
isomorphism of the bundle $T\oplus T^{\ast}$ and also maps the triple
$({\cal J}_{J},{\cal J}_{\omega},G)$ in a well defined way. This means that 
on the mirror side this triple is completely fixed.

Let the mirror map ${\cal M}:T\oplus T^{\ast}\to T\oplus T^{\ast}$ be given by
\be510
{\cal M}=\begin{pmatrix} 1 &  &  &  \\
         &  &  & 1 \\
         &  & 1 &  \\
         & 1 &  &  
\end{pmatrix}
\ee
where we distinguish the vielbeins (as above) in $T$ and $T^{\ast}$, more
precisely, 
\be511
{\cal M}:T_{ {\cal B} }\oplus T_{ {\cal F} }\oplus T_{ {\cal B} }^{\ast}\oplus
T_{ {\cal F} }^{\ast} \to
T_{ {\cal B} }\oplus T^{\ast}_{ {\cal F} }\oplus T_{ {\cal B} }^{\ast}\oplus
T_{ {\cal F} }
\ee
This is
simply a map acting on the base as the identity and on the fibre as a
``flip''. Note: The identity maps in ${\cal M}$ are tensors of 
adequate type to make ${\cal M}$ a well defined isomorphism. The property
${\cal M}={\cal M}^{-1}$ makes the mirror map an involution, ${\cal M}^2=1$. 

The action of the mirror map on a generalized structure 
$({\cal J}_{J},{\cal J}_{\omega},G)$ is defined by,
\be514
(\hat {\cal J}_{J},\hat {\cal J}_{\omega},\hat G)=
   {\cal M} \, ({\cal J}_{J},{\cal J}_{\omega},G) \, {\cal M}^{-1} \, .
\ee 

Let us act now by ${\cal M}$ on the generalized structures defining the $T^6$.
We obtain the mirror metric $\hat G$ by (because of ${\cal M}={\cal M}^{-1}$
we abuse notation and write only ${\cal M}$), 
\be515
\hat G = {\cal M} \, G \, {\cal M}
=\begin{pmatrix}  &  & \delta^{ij} &  \\
         &  &  & \delta_{\alpha\beta} \\
         \delta_{ij}&  &  &  \\
         & \delta^{\alpha\beta} &  &  
\end{pmatrix} \, .
\ee
The mirror metric $\hat g$ on the tangent bundle is therefore given by
\be520
\hat g= \begin{pmatrix} \delta_{ij} &  \\ 
                 & \delta^{\alpha\beta}\end{pmatrix} \, ,
\ee
where we do have now the inverse metric in the trivial fibre
which is expected by using the Buscher rules. By a similar mirror 
transformation of $({\cal J}_{J},{\cal J}_{\omega})$ we get
\be525
(\gencm,\gensm) = (\M\genc\M,\M\gens\M) \, ,
\ee
where we obtain
\be530
\gencm=\begin{pmatrix}  &  & &  -\delta^{\alpha i} \\
         &  & \delta^{i\alpha} &  \\
         & -\delta_{\alpha i} &  &  \\
         \delta_{i\alpha}&  &  &  
\end{pmatrix}, 
\qquad
\gensm=\begin{pmatrix}  & -\delta_{\alpha}{}^i & &  \\
         \delta_i{}^{\alpha}&  &  &  \\
         &  &  & -\delta^{\alpha}{}_i  \\
         &  & \delta^i{}_{\alpha} &  
\end{pmatrix} \, .
\ee
Considering these structures as an embedding of usual complex and symplectic
structures (acting in the tangent bundle) we see immediately that 
$\gencm$ is of pure
symplectic type ($k=0$) while $\gensm$ is of pure complex type ($k=3$). 
This makes the mirrored structures of generalized type and 
agrees with the literature (see e.g. \cite{240}\cite{210}) that on the mirror 
space the (generalized) algebraic structures are interchanged:
\renewcommand{\arraystretch}{1.8}
\be531
\ba{rcl}
\varphi_1=e^{i\, \omega} & \leftrightarrow & \hat\varphi_1=\Omega\\
\varphi_2=\Omega & \leftrightarrow & \hat\varphi_2=e^{i\, \omega}\\
g_{ {\cal B}} + g_{ {\cal F}} & \leftrightarrow & 
                  g_{ {\cal B}} + g^{-1}_{ {\cal F}}\\
\mbox{trivial GCY} & \leftrightarrow & \mbox{trivial GCY} \, .
\renewcommand{\arraystretch}{1.2}
\ea
\ee
We verify (for $T^6$) therefore the work of \cite{240} (see also \cite{210}).
But there the authors must introduce the Buscher rules by hand, 
in contrast to the mappings above, where these rules are already included.
%
\subsection{$B$-field transform of $T^6$}
%
Next in our line is the introduction of a globally defined and flat $B$-field.
We are considering only closed $B$-fields because such a transformation
of a pure spinor will not affect the integrability. We only single out 
different involutive $C_{\pm}$. 

We do not consider general $B$-fields within this article. Let 
$B= B_{i\alpha} dy^i\wedge dx^{\alpha}$ be the $B$-transformation
of our interest:
\be550
e^B=\begin{pmatrix} 1 & & &  \\
                    & 1 &  &  \\
                    & B_{\alpha i}  & 1 &   \\
              B_{i\alpha}      &  &  & 1 
\end{pmatrix},
\qquad
e^{-B}=\begin{pmatrix} 1 & & &  \\
                    & 1 &  &  \\
                    & -B_{\alpha i}  & 1 &   \\
              -B_{i\alpha}      &  &  & 1 
\end{pmatrix}
\ee 
The $B$-transformed metric $G$ of the $T^6$ can be computed by
\be555
G^{B} = e^B \, G \, e^{-B}
\ee
but we suppress here the explicit form. By a following mirror transformation
of this metric we obtain 
\be560
\hat G^{B} = \M G^{B} \M,
\ee
which is completely off-diagonal. It is not hard to proof that it is again
a generalized metric and more important of pure Riemannian type. 
Or in other words $\hat G^{B}$ is a ``pure'' Riemannian metric (not
$B$-transformed) and thus can 
be constructed by embedding of only a Riemannian metric $\hat g$,
\be565
\hat g= \begin{pmatrix} 
     g_{ {\cal B}} -Bg_{ {\cal F}}^{-1}B & Bg_{ {\cal F}}^{-1} \\ 
                -g_{ {\cal F}}^{-1}B & g_{ {\cal F}}^{-1}
\end{pmatrix} 
\ee
where $g_{ {\cal B}}=\delta_{ij}$ is the metric in the base ${\cal B}$ and
$g_{ {\cal F}}=\delta_{\alpha\beta}$ denotes the metric in the fibre 
${\cal F}$. These transformation rules do verify the Buscher rules exactly
and means that on the mirror side there is no $\hat B$-field and the ``old''
one is completely absorbed in the metric $\hat G$.

\begin{example} Let $B=(B_{i\alpha}dy^i)\wedge dx^{\alpha}$ (in local
coordinates) only depends on coordinates $y$ on the base. The mirror metric
$\hat g$ has the shape of the following form
\be566
\hat g = (g_{ {\cal B}})_{ij}dy^idy^j + (g_{ {\cal F}}^{-1})_{\alpha\beta}
       (dx^{\alpha} + A^{\alpha})(dx^{\beta} + A^{\beta})
\ee
where $A$ is a local, flat connection-one-form. Thus, the $B$-field
transforms into the metric and re-appears in a non-trivial fibration.
\end{example}
Moreover, because of ${\cal M}$ being an isomorphism and an involution we
can reverse the procedure. The initial data, a flat non-trivial fibration
and vanishing $B$-field, mirrors in a trivial fibred $T^6$ with 
non-trivial $B$-field. 
Naturally, the combination of both ``effects'' is possible and
independent of each other.

The $B$-field transformation of the generalized complex structures 
$(\genc,\gens)$ can be calculated by
\be570
(\gencb,\gensb)=(e^B \genc e^{-B} , e^B \gens e^{-B})
\ee
which are no longer diagonal/off-diagonal, respectively. 
An action of the mirror map $\M$ on these is given by
\be575
(\gencm,\gensm)= (\M\gencb\M,\M\gensb\M).
\ee 
It can also be shown
that the mirror structures are generalized complex structures and
become again completely 
off-diagonal/diagonal, or equivalently, 
structures of pure symplectic and complex type, respectively 
(see also \cite{210}). 
The embedded structures are given in components by
\be580
\hat J= \begin{pmatrix} 
      \delta B & -\delta \\ 
       B\delta^{-1} B +\delta & -B\delta^{-1}
\end{pmatrix},
\qquad
\hat \omega= \begin{pmatrix} 
      B\delta+\delta B & -\delta \\ 
      \delta  & 0
\end{pmatrix}
\ee
We used a condensed notation where $\delta$ denotes the appropriate tensors
(see(\ref{305}),(\ref{306})) and we also suppressed here the identity maps
coming from ${\cal M}$. Note: The object $\hat J$ is embedded in 
$\gensm$ (which makes it a complex structure (type $k=3$)) and should 
therefore not mixed up with the subscript $\omega$ which denotes the
symplectic form on the $T^6$ by which we started with.
%
\section{Generalizations}
%
The concepts that where introduced in the previous section 
(see also \cite{210}) hold in more generality, and
we only applied it to a simple example to make the mappings of the
generalized structures clear. So one can immediately use the framework for
more complicated generalized (almost) Calabi-Yau structures, being
torsion-full and non-integrable in general. For example if the base is no 
longer $T^3$ and is 
a more general 3-dimensional manifold. Naturally, some base allow for 
non-trivial fibrations of $T^3$ (or even more general (S)LAG fibrations). 
The $B$-field can be treated in more general sense as well by dropping the
restrictions and moreover the flatness assumption. 

The previous discussion means we have to discuss integrability (and torsion) 
in more detail. Focusing 
on the above considered cases, obviously, these are integrable 
and torsion-less and so is its mirror. The case of $NS$-fluxes 
is already worked out in the amazing paper of 
\cite{240} where the authors also used the concept of generalized
complex structures but they had to introduce the Buscher rules by hand.
There it was shown how torsion-full 6-manifolds
are interchanged, and the authors gave a precise description of each $SU(3)$ 
torsion component. So it must be possible to recalculate these results by the
concepts given in this article and by dropping the above assumed 
restrictions (made only for convenience). 

Relating the introduced notation with that in physics makes it immediately
clear that the above $B$-field is defined globally. But the
physical meaning of a $B$-field is actually different. There a $B$-field
is only defined patchwise and serves as a connection of a $1$-gerbe $H$
which takes values in $H^3(M^6,{\mathbb Z})$ (see also \cite{180}). 
In the language of branes, this corresponds to topologically
inequivalent embeddings of branes carrying $NS$-flux. Furthermore, if also
D-branes, SUSY-compatible submanifolds carrying vector-bundles charged under
K-theory, are present, the $NS$-flux serves as a twist. This means that
it is more close to physics to bring the physical $B$-field via the
twisted differential operator (\ref{255}) into play. In the case of $T^6$ 
it becomes immediately clear why it is not possible to have $dB=H\neq 0$.
Using once more generalized structures, D-branes should (actually) be treated
like generalized submanifolds(see e.g. 
\cite{150,160,190,200,210} for definitions 
and recent developments). 

\section*{Acknowledgments}

The author would thank F. Gmeiner, D. L\"ust and  
S. Stieberger for very valuable discussions as well as 
J. Babington and I. Runkel for proofreading the draft version. Special
thanks go to F. Witt for very usefull correspondence, mathematical support 
and proofreading.

The work of C.J. is supported by a Graduiertenkolleg grant of the DFG 
(The Standard Model of Particle
Physics - structure, precision tests and extensions).



\begin{thebibliography}{10}

\bibitem{140}
A.~Strominger, S.-T. Yau, and E.~Zaslow, ``Mirror symmetry is {T}-duality,''
  {\em Nucl. Phys.} {\bf B479} (1996) 243--259,
\href{http://www.arXiv.org/abs/hep-th/9606040}{{\tt hep-th/9606040}}.

\bibitem{185}
N.~Hitchin, ``Generalized {C}alabi-{Y}au manifolds,'' {\em Q. J. Math.} {\bf 54
  no.3} (2003) 281--308,
\href{http://www.arXiv.org/abs/math.dg/0209099}{{\tt math.dg/0209099}}.

\bibitem{190}
M.~Gualtieri, {\em Generalized complex geometry}.
\newblock PhD thesis, Oxford University, 2004.
\newblock
\href{http://www.arXiv.org/abs/math.dg/0401221}{{\tt math.dg/0401221}}.
\newblock

\bibitem{110}
U.~Lindstrom, R.~Minasian, A.~Tomasiello, and M.~Zabzine, ``Generalized complex
  manifolds and supersymmetry,''
\href{http://www.arXiv.org/abs/hep-th/0405085}{{\tt hep-th/0405085}}.

\bibitem{150}
A.~Kapustin, ``Topological strings on noncommutative manifolds,''
\href{http://www.arXiv.org/abs/hep-th/0310057}{{\tt hep-th/0310057}}.

\bibitem{160}
M.~Zabzine, ``Geometry of {D}-branes for general {$N=(2,2)$} sigma models,''
\href{http://www.arXiv.org/abs/hep-th/0405240}{{\tt hep-th/0405240}}.

\bibitem{180}
N.~Hitchin, ``Lectures on special {L}agrangian submanifolds,'' in {\em {W}inter
  {S}chool on {M}irror {S}ymmetry, {V}ector {B}undles and {L}agrangian
  {S}ubmanifolds {(}{C}ambridge, {MA}, 1999{)}}, A.~{S}tud.~{A}dv. {M}ath.,
  ed., vol.~23, pp.~151--182.
\newblock Amer. Math. Soc., Providence, RI, 2001.
\newblock
\href{http://www.arXiv.org/abs/math.dg/9907034}{{\tt math.dg/9907034}}.
\newblock

\bibitem{200}
O.~Ben-Bassat and M.~Boyarchenko, ``Submanifolds of generalized complex
  manifolds,''
\href{http://www.arXiv.org/abs/math.dg/0309013}{{\tt math.dg/0309013}}.

\bibitem{210}
O.~Ben-Bassat, ``Mirror symmetry and generalized complex manifolds,''
\href{http://www.arXiv.org/abs/math.ag/0405303}{{\tt math.ag/0405303}}.

\bibitem{220}
S.~F. Hassan, ``{$O(d,d:R)$} deformations of complex structures and extended
  world sheet supersymmetry,'' {\em Nucl. Phys.} {\bf B454} (1995) 86--102,
\href{http://www.arXiv.org/abs/hep-th/9408060}{{\tt hep-th/9408060}}.

\bibitem{230}
S.~F. Hassan, ``{$SO(d,d)$} transformations of {R}amond-{R}amond fields and
  space-time spinors,'' {\em Nucl. Phys.} {\bf B583} (2000) 431--453,
\href{http://www.arXiv.org/abs/hep-th/9912236}{{\tt hep-th/9912236}}.

\bibitem{240}
S.~Fidanza, R.~Minasian, and A.~Tomasiello, ``Mirror symmetric
  {$SU(3)$}-structure manifolds with {NS} fluxes,''
\href{http://www.arXiv.org/abs/hep-th/0311122}{{\tt hep-th/0311122}}.

\end{thebibliography}

\providecommand{\href}[2]{#2}\begingroup\raggedright\endgroup

\end{document}